\documentclass[runningheads]{llncs}
\usepackage{graphicx}
\usepackage{amsmath,amssymb,amsfonts}
\usepackage{hyperref}
\usepackage[english]{babel}
\usepackage{subcaption}
\usepackage{mwe}
\usepackage[T1]{fontenc}
\usepackage{lmodern}
\newtheorem{observation}[theorem]{\textbf{Observation}}

\def\mi#1{\mathit{#1}}
\usepackage{amsmath}
\def\mc#1{\mathcal{#1}}
\usepackage[font=small,labelfont=bf]{caption}

\def\mi#1{\mathit{#1}}
\def\mc#1{\mathcal{#1}}
\usepackage[activate={true,nocompatibility},final,tracking=true,kerning=true,spacing=true,factor=1100,stretch=10,shrink=10]{microtype}

\addto\extrasenglish{%
}

\begin{document}
\title{Discovering High-Quality Process Models Despite Data Scarcity}
%
%

\author{Jan Niklas Adams\inst{1}\and
Jari Peeperkorn\inst{2}\and
Tobias Brockhoff\inst{1}\and
Isabelle Terrier\inst{3}\and
Heiko Göhner\inst{3}\and
Merih Seran Uysal\inst{1}\and
Seppe vanden Broucke\inst{4,2}\and
Jochen De Weerdt\inst{2}\and
Wil M.P. van der Aalst\inst{1}
}
\authorrunning{J.N. Adams et al.}
%
\institute{Chair of Process and Data Science, RWTH Aachen University, Aachen, Germany \\
\email{\{niklas.adams,brockhoff,uysal,wvdaalst\}@pads.rwth-aachen.de}
\and
Research Center for Information Systems Engineering (LIRIS), KU Leuven
\email{\{jari.peeperkorn,seppe.vandenbroucke,jochen.deweerdt\}@kuleuven.be}
\and
Heidelberger Druckmaschinen AG, Heidelberg, Germany
\and Department of Business Informatics and Operations Management, Ghent University
\\}
\maketitle              
%
\begin{abstract}
Process discovery algorithms learn process models from executed activity sequences, describing concurrency, causality, and conflict. 
Concurrent activities require observing multiple permutations, increasing data requirements, especially for processes with concurrent subprocesses such as hierarchical, composite, or distributed processes.
While process discovery algorithms traditionally use sequences of activities as input, recently introduced object-centric process discovery algorithms can use graphs of activities as input, encoding partial orders between activities. As such, they contain the concurrency information of many sequences in a single graph.
In this paper, we address the research question of reducing process discovery data requirements when using object-centric event logs for process discovery.
We classify different real-life processes according to the control-flow complexity within and between subprocesses and introduce an evaluation framework to assess process discovery algorithm quality of traditional and object-centric process discovery based on the sample size. We complement this with a large-scale production process case study. 
Our results show reduced data requirements, enabling the discovery of large, concurrent processes such as manufacturing with little data, previously infeasible with traditional process discovery. Our findings suggest that object-centric process mining could revolutionize process discovery in various sectors, including manufacturing and supply chains.
\end{abstract}

\section{Introduction}
Throughout time, business endeavors have been supported by processes, ranging from bookkeeping to production processes. With the advent of modern information systems, data traces of process executions are recorded in the underlying databases~\cite{DBLP:books/sp/DumasRMR18}. While businesses mostly have some idea of how their processes look like, analyzing the recorded data allows them to uncover their real execution. Algorithms transforming the data of process executions, i.e., \textit{event logs}, into process models are called \textit{process discovery} algorithms~\cite{DBLP:books/sp/Aalst16}.

A plethora of process discovery algorithms have been proposed over the last two decades~\cite{DBLP:journals/tkde/AugustoCDRMMMS19}. Process discovery techniques assume the existence of an extracted event log composed of a set of event sequences (\textit{cases}) describing different process executions. In general, these algorithms project each event sequence to its \textit{activity} sequence and aim to uncover the sequentiality/causality, concurrency, or conflict between activities. While causality and conflict can be uncovered using relatively few activity sequences, concurrency needs to be confirmed using a large set of observations. For example, four concurrent activities could manifest in 24 different sequences, and seven concurrent activities already in 5040 possible sequences. The number of sequences grows factorial with the number of concurrent activities. Therefore, larger numbers of concurrent activities require more observations for process discovery. This becomes infeasible for large numbers of concurrent activities or event logs with very few cases.

\begin{figure}[t]
    \centering
    \includegraphics[width = 0.7\textwidth]{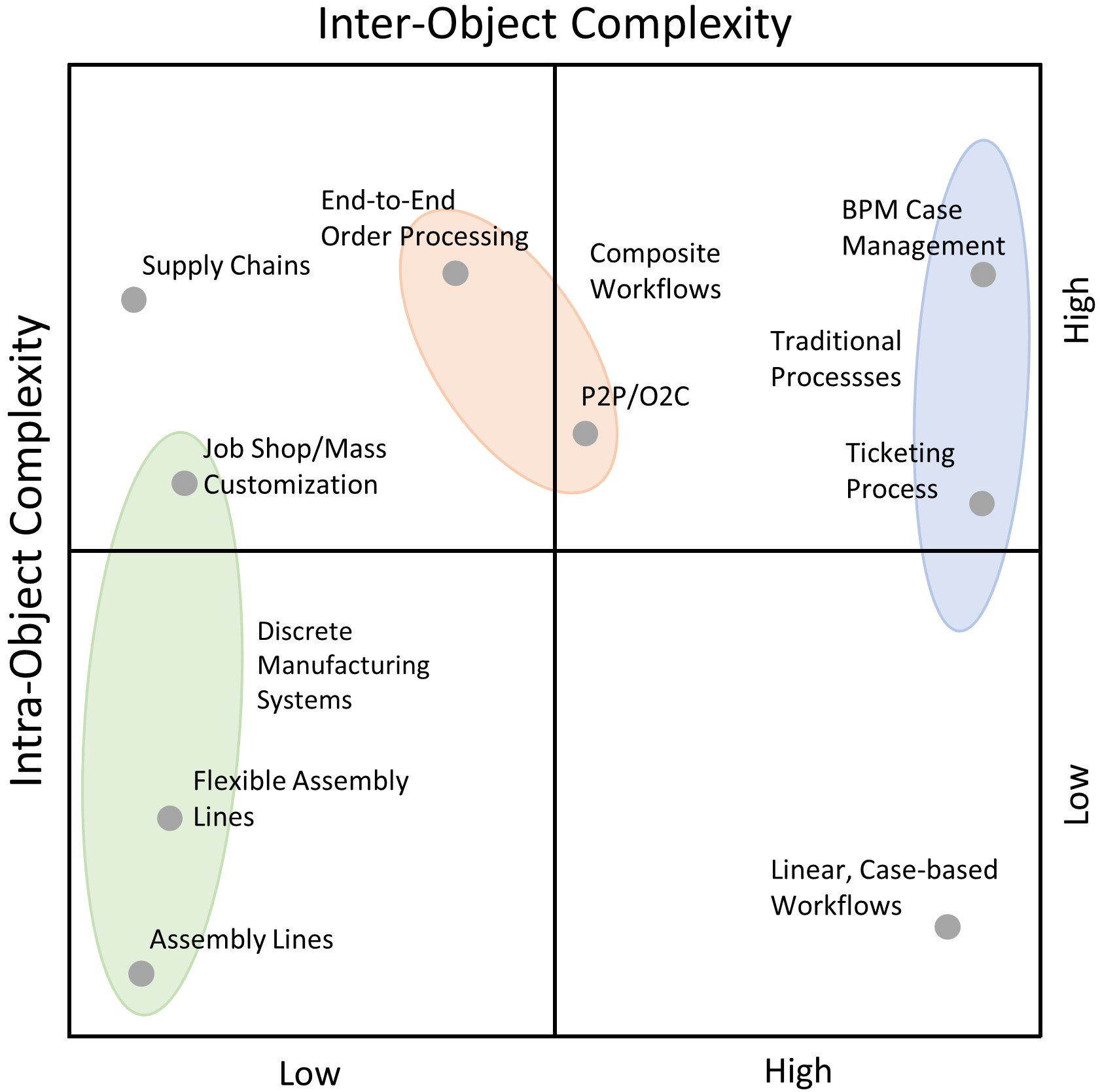}
    \caption{Categorization of different real-life processes based on inter- and intra-object complexity.}
    \label{fig:taxonomy}
\end{figure}

Problems with highly concurrent behavior are especially present in processes that are composed of multiple, concurrently running and lightly interacting subprocesses. The amount of interaction between subprocesses is called the \textit{inter-object complexity}. High inter-object complexity indicates a tight coupling between subprocesses, i.e., large amounts of shared control flow. Low inter-object complexity indicates concurrently running subprocesses. We depict an overview of typical processes and their inter-object complexity in \autoref{fig:taxonomy}. As inter-object complexity only captures control-flow complexity between subprocesses, we complement this dimension with \textit{intra-object complexity}, capturing the control-flow complexity within subprocesses. We collect descriptions of typical real-life processes from the literature and depict their mapping onto these two dimensions in \autoref{fig:taxonomy}. Specifically, we show processes considered in traditional process mining (workflows like ticketing processes and business process case management~\cite{DBLP:journals/dke/AalstDHMSW03,DBLP:books/sp/22/WeerdtW22}), discrete manufacturing systems~\cite{KIM19933} (job shop/mass customization~\cite{DASILVEIRA20011}, (flexible) assembly lines~\cite{KIM19933}), and composite workflows (business process like P2P/O2C~\cite{DBLP:conf/kdd/ZengMLBM08}, end-to-end order processing~\cite{DBLP:conf/ieem/SchuhGCSA20}), and supply chains~\cite{milgate2001supply}. Additionally, we depict where a traditional, completely linear process would be positioned.

While traditional event logs record cases as sequences of events, \textit{Object-Centric Event Logs} (OCELs)~\cite{DBLP:conf/sefm/Aalst19} encode cases as graphs of events~\cite{ADAMS_ICPM_2022}, preserving partial orders between events. By extracting the event data of compound processes as OCELs instead of traditional event logs, the concurrency between the subprocesses can be preserved using these graph-based process executions. Object-centric process discovery was recently introduced~\cite{DBLP:journals/fuin/AalstB20}, using an OCEL as input and discovering a model composed of multiple interacting subprocesses. This enables process owners to discover a process directly from the graph-based process executions, rather than forcing them into the sequential structure needed for traditional process discovery and removing concurrency information.

In this paper, we address the following research question: How do data requirements reduce when using object-centric process discovery instead of traditional process discovery? Specifically, we want to break down the effect of employing object-centric discovery for different groups of processes with respect to their control-flow complexity between and within subprocesses, mapping the reduction in data requirements back to real-life processes.

We tackle these research questions through two main contributions: First, we formally define inter- and intra-object complexity and an evaluation framework that can exactly assess the discovered model quality based on the sample size of an event log. We instantiate this evaluation framework with ~45.000 model and sample size combinations and compare the quality of the discovered model between traditional discovery and object-centric discovery. We map the results back to the dimension of inter- and inter-object complexity and assess which processes have the highest data requirement reduction when using object-centric process discovery. Second, we complement this experimental evaluation with a large-scale case study. Since the experimental evaluation is very computation heavy, it is only applicable to smaller models. Our case study shows the successful application of object-centric process discovery on a large real-life manufacturing process with hundreds of activities. It also shows that traditional process discovery fails to deliver the same quality.

The remainder of this paper is structured as follows. We introduce the related work on process discovery and object-centric process mining in \autoref{sec:rel_work}. The preliminaries on traditional and OCELs and discovery are presented in \autoref{sec:prelim}. We propose an experimental framework comparing discovery on object-centric and flattened event data in \autoref{sec:method}. We discuss these results in \autoref{sec:results}. A case study for the advantages of object-centric discovery is presented in \autoref{sec:casestudy}. We conclude this paper in \autoref{sec:conclusion}.

\section{Related Work}
\label{sec:rel_work}
Object-centric process mining addresses the problem of distinguishing multiple objects involved in a process. These objects can be used to identify subprocesses. In the past, this problem has extensively been studied from the modeling side. Different process modeling languages/notations to capture object-centric processes have been proposed, such as artifact-centric process models~\cite{DBLP:journals/debu/CohnH09}, \textit{Object-Centric Behavioral Constraints} (OCBC)~\cite{DBLP:conf/dlog/AalstAMT17}, proclets~\cite{DBLP:conf/apn/Fahland19}, DB-Nets~\cite{DBLP:journals/topnoc/MontaliR17}, COA-Nets~\cite{DBLP:conf/bpm/GhilardiGMR20}, and t-PNIDs~\cite{DBLP:conf/apn/WerfRPM22}. Some of these modeling techniques have been accompanied by a data format that is able to capture event data for processes of this kind. Artifact-centric event logs~\cite{DBLP:conf/bpm/NooijenDF12,DBLP:conf/IEEEscc/Moctar-MBabaASG22}, and \textit{eXtensible Object-Centric event logs} XOC~\cite{DBLP:conf/caise/LiMCA18} have been proposed. Furthermore, Esser and Fahland have recently proposed a general-purpose graph database to store object-centric event data~\cite{DBLP:journals/jodsn/EsserF21}. Object-centric process mining tackles the problem of object-centricity with the modeling language of object-centric Petri nets~\cite{DBLP:journals/fuin/AalstB20} and the accompanying data format of OCELs~\cite{DBLP:conf/sefm/Aalst19}. Both of these approaches have been developed to improve the complexity and scalability issues of existing modeling and data storage formats. 

The core motivation of process discovery is the uncovering of as-is processes from real-life data. As such, process-discovery techniques aim to connect event data with process models. For the case of object-centric process mining, discovery techniques have been proposed for the modeling languages where data storage formats exist, namely artifact-centric discovery~\cite{DBLP:conf/bpm/NooijenDF12} and OCBC discovery from XOC~\cite{DBLP:conf/bis/LiCA17}. Van der Aalst and Berti have introduced a general approach for discovering object-centric Petri nets from object-centric event data~\cite{DBLP:journals/fuin/AalstB20}. Subsequently, these process models are utilized in other data-driven process mining tasks, such as conformance checking~\cite{DBLP:conf/icpm/AdamsA21_new} or enhancement~\cite{DBLP:conf/er/ParkAA22_new}. This approach applies a traditional process discovery to each object type and merges the resulting models at interaction points. 
Any traditional process discovery algorithm can be employed in object-centric discovery. 

Traditional process discovery describes the problem of mapping a set of activity sequences to a process model~\cite{DBLP:books/sp/Aalst16}. Many different techniques have been proposed~\cite{DBLP:journals/tkde/AugustoCDRMMMS19}, where the Split Miner~\cite{DBLP:journals/kais/AugustoCDRP19} and the Inductive Miner~\cite{DBLP:conf/apn/LeemansFA13} remain among the most popular algorithms. In our paper, we use the inductive miner due to its guarantee of sound workflow nets for every individual object type.  This paper provides a quantitative and qualitative study as to how much process discovery results can be improved by individually discovering and merging process models for each object type rather than discovering one model on the flattened event data of all object types.

\section{Traditional and Object-Centric Process Discovery}
\label{sec:prelim}
We introduce the foundations of object-centric and traditional event data and process discovery in this section. We link OCELs to traditional event logs and define the flattening of a traditional event log. Analogously, we introduce object-centric Petri nets and their relationship to traditional Petri nets.

\subsection{Linking Object-Centric and Traditional Event Data}
First, we introduce some notations used throughout this paper.
$\mc{E}$ is the universe of event identifiers,
$\mc{OT}$ is the universe of object types, and
$\mc{O}$ is the universe of objects. Each object is associated with exactly one object type through $\pi_\mi{type}:\mc{O}\rightarrow \mc{OT}$.
$\mc{T}$ denotes the universe of event timestamps and
$\mc{A}$ the universe of event activities.
We denote the powerset of a set $X$, the set of all subsets, with $\mc{P}(X)$. A sequence of length $n\in\mathbb{N}$ orders elements of a set $X$ and is denoted with $\sigma : \{1,\ldots , n\}\rightarrow X$ and $\sigma = \langle x_1,\ldots, x_n \rangle$. The set of directly follows relationships in a sequence is denoted by $\mi{df}(\langle x_1,\ldots, x_n \rangle) = \{(x_i,x_i+1)\mid i\in\{1,\ldots, n-1\}\}$.

\begin{definition}[Object-Centric Event Log]
An object-centric event log is a tuple $L=(E, \allowbreak O \allowbreak ,OT,\pi_\mi{act},\pi_\mi{time},\pi_\mi{trace})$ consisting of 
\begin{itemize}
    \item[$\bullet$] events $E\subseteq \mc{E}$, objects $O\subseteq \mc{O}$, object types $OT = \{\pi_\mi{type}(o) \mid o \in O \}$,
    \item[$\bullet$] activities $\pi_\mi{act}:E\rightarrow \mc{A}$,  timestamps $\pi_\mi{time}:E\rightarrow \mc{T}$, and
    \item[$\bullet$] ordering $\pi_\mi{trace}:O\rightarrow E^*$ mapping each object to a sequence of events such that $\forall_{o \in O}\; \pi_\mi{trace}(o) = \langle e_1,\ldots, e_n \rangle \wedge \forall_{i\in \{1,\ldots, n-1\}}\; \pi_\mi{time}(e_i) \leq \pi_\mi{time}(e_{i+1})$
    
\end{itemize}
Each event is linked to objects $\pi_\mi{obj}(e) = \{o\in O\mid e\in \pi_\mi{trace}(o) \}$.
    
\end{definition}
\begin{figure}[t]
    \centering
    \includegraphics[width = 0.8\textwidth]{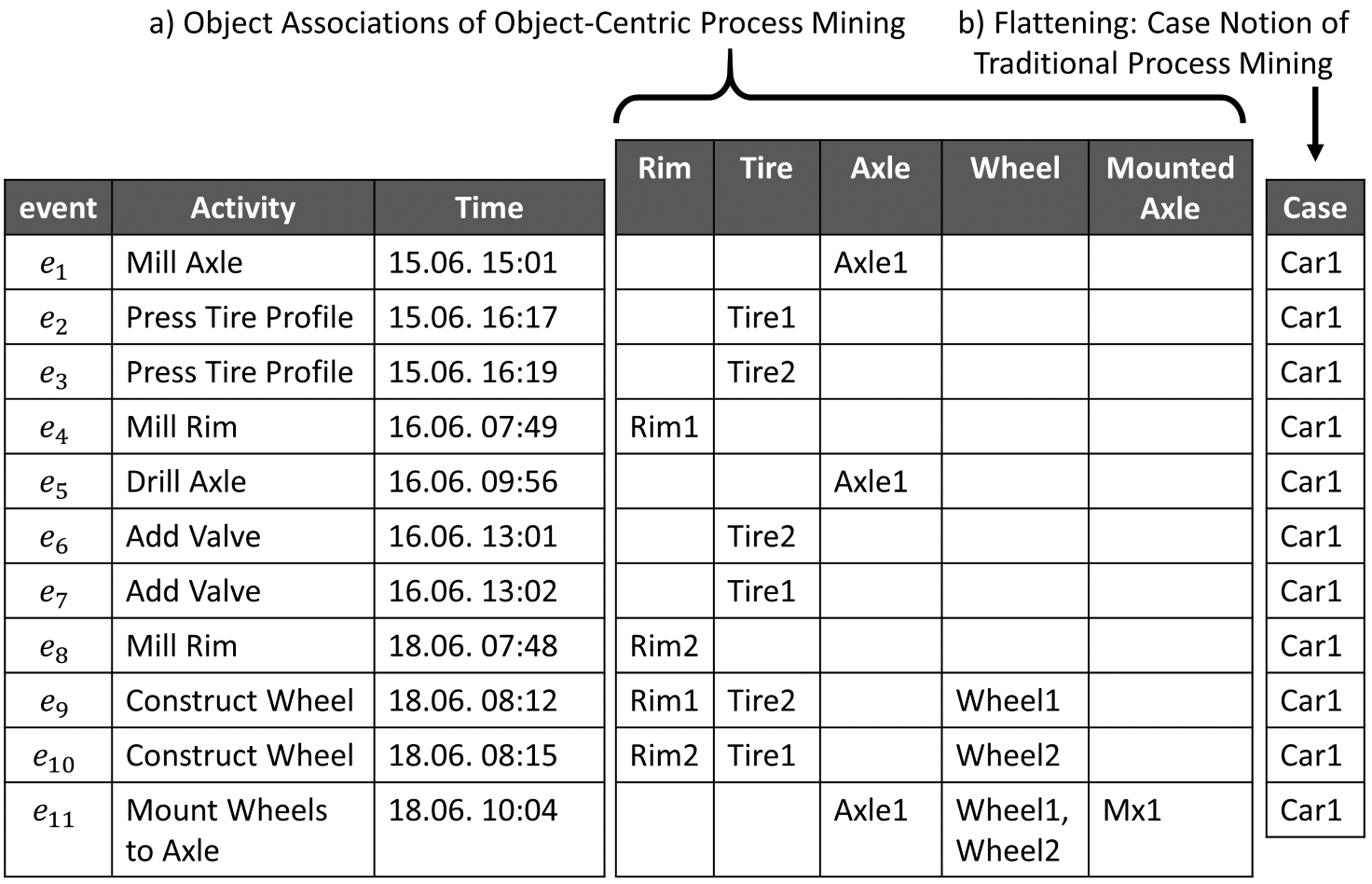}
    \caption{An object-centric event log (a) and its flattened counterpart (b).}
    \label{fig:ocel}
\end{figure}
An example of an OCEL is depicted in \autoref{fig:ocel}. Each row is an event associated with objects of different types, an activity, and a timestamp. Each object is associated with a sequence of events, e.g., $\pi_\mi{trace}(\mi{Tire2}) = \langle e_2, e_7, e_{10} \rangle$.

A traditional event log is a special case of an OCEL where objects are all of the same type and an event is associated with exactly one object.
\begin{definition}[Traditional Event Log]
An object-centric event log $L{=}(E, \allowbreak O, \allowbreak OT,\pi_\mi{act},\pi_\mi{time},\pi_\mi{trace})$ is a traditional event log iff $|OT| {=} 1 \wedge  \forall_{e\in E} |\pi_\mi{obj}(e)| {=} 1$.
\end{definition}

An event log consists of different executions of the same process. Each execution contains multiple events. In traditional process mining, a process execution is called a \textit{case} and is a sequence of events for one object. We generalize this notion such that a process execution is a graph of events for multiple, connected objects~\cite{ADAMS_ICPM_2022}. The graph describes a partial order of events induced by the precedence constraints of each $\pi_\mi{trace}$ of the involved objects.
\begin{definition}[Process Executions~\cite{ADAMS_ICPM_2022}]
Let $L=(E, \allowbreak O \allowbreak ,OT,\pi_\mi{act},\pi_\mi{time},\pi_\mi{trace})$ be an object-centric event log. The object graph of object co-appearances is defined by $\mi{OG}_L=(O,I)$ with $I = \{\{o,o'\}\mid o\neq o' \wedge  \exists_{e\in E}\; \{o,o'\}\subseteq \pi_\mi{obj}(e) \}$. The set of interdependent object sets are the connected components of the object graph $\mi{cc}(L) =\{O' \subseteq O \mid O' \text{ form a connected component in } \mi{OG}_L\}$. One set of dependent objects $X\in \mi{cc}(L)$ spans a process execution. A process execution is a graph $p_X = (E_X, D_X)$ of nodes $E_X = \{e\in E\mid \pi_\mi{obj}(e) \cap X \neq \emptyset \}$ and edges $D_X = \{ (e,e') \in E_X \times E_X \mid \exists_{o\in X}\; ( e,e') \in \mathit{df}(\pi_\mi{trace}(o))\}$. The set of all process executions of an event log is defined by $\mi{px}(L)=\{p_X\mid X \in \mi{cc}(L)\}$.
\end{definition}
\begin{figure}[t]
    \centering
    \includegraphics[width = \textwidth]{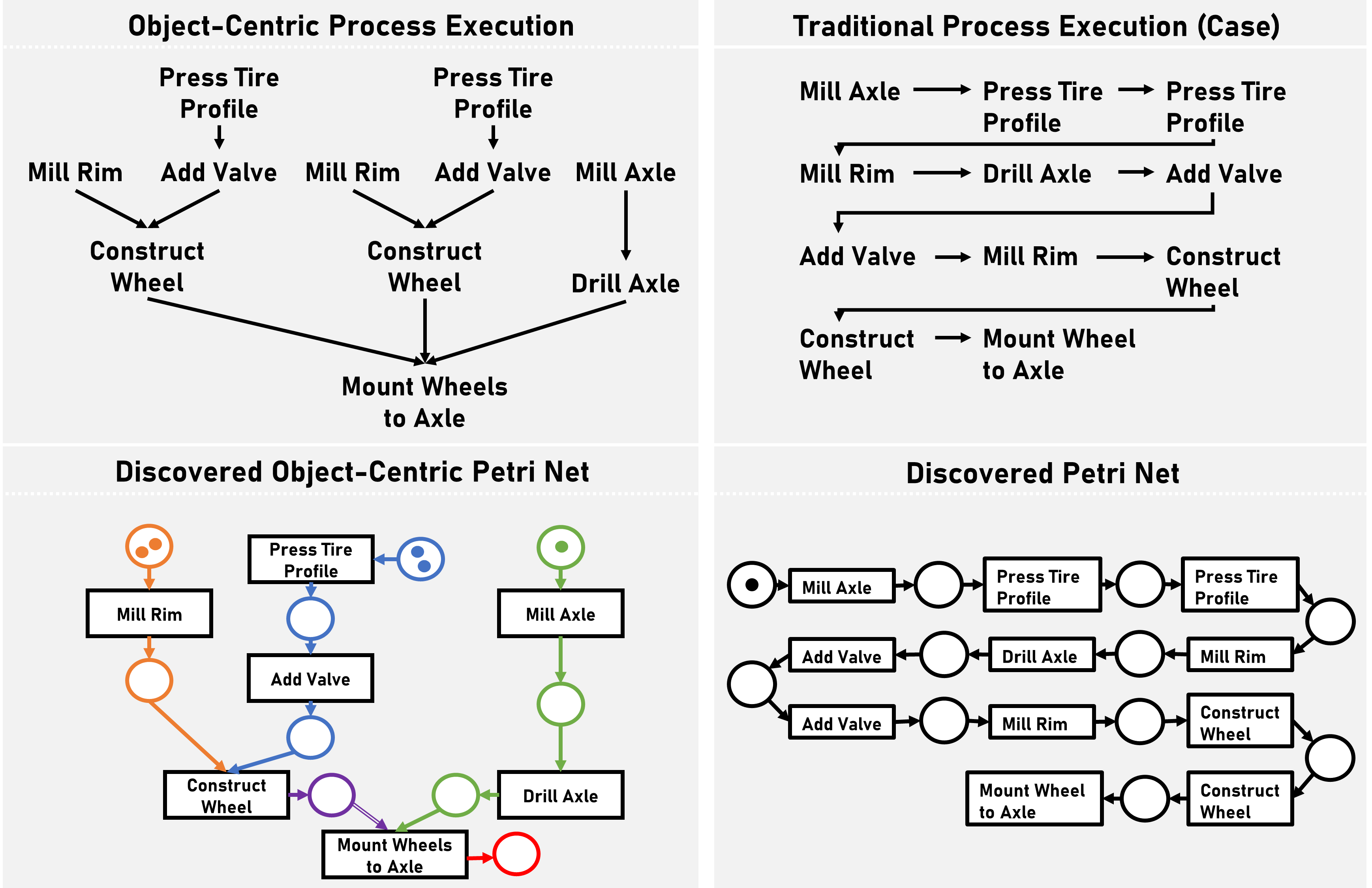}
    \caption{Top: A graph-based process execution from the object-centric event log of \autoref{fig:ocel} a) and the sequential process execution from the flattened, traditional event log of \autoref{fig:ocel} b). Bottom: The process models discovered from both of these process executions.}
    \label{fig:p_exec}
\end{figure}
The upper left part of \autoref{fig:p_exec} contains an example of the object-centric process execution contained in \autoref{fig:ocel}. The process execution shows the production of different parts that run concurrently and are assembled in a bottom-up way.

An OCEL can be transformed into the traditional event log format by \textit{flattening}~\cite{DBLP:conf/sefm/Aalst19} it. This involves the choice of a case notion and the sequentializing of events for each object of that case notion. The case notion can be chosen freely, typically either a single object type is chosen or a set of connected objects. Flattening transforms a graph-based structure of process executions into a less expressive sequential structure of process execution, therefore, some information loss will never be preventable. We choose to flatten with all connected objects, i.e., flattening a process execution from a graph into a sequence. By doing this, we prevent events from disappearing or being duplicated as it would be the case if one would flatten on a single object type~\cite{DBLP:conf/sefm/Aalst19}.
\begin{definition}[Flattening]
Let $L=(E, \allowbreak O \allowbreak ,OT,\pi_\mi{act},\pi_\mi{time},\pi_\mi{trace})$ be an object-centric event log.
$L^\mi{flat}=(E, \allowbreak O' \allowbreak ,OT',\pi_\mi{act},\pi_\mi{time},\pi_\mi{trace}')$ is the flattened event log with three modified elements:
\begin{itemize}
    \item[$\bullet$] a new, single object type $OT' = \{ot\}$ for an $ot\in \mc{OT} \wedge ot\notin OT$.
    \item[$\bullet$] new object identifiers $O' \subseteq \{o\in \mc{O}\mid \pi_\mi{type}(o) \in OT'\}$ associated with the objects of the process executions $\mi{cc}(L)$ through a bijection $\pi_\mi{flat}:O' \rightarrow \mi{cc}(L)$.
    \item[$\bullet$] event sequences for the new objects $\pi_\mi{trace}'(o') = \langle e_1, \ldots, e_n \rangle$ with $\{e_1,\ldots, e_n\} = \bigcup_{o \in \pi_\mi{flat}(o')} \pi_\mi{trace}(o)$ and $\pi_\mi{time}(e_1) \leq \cdots \leq \pi_\mi{time}(e_n)$ for $o' \in O'$.
\end{itemize}
\end{definition}
We show the flattened event log in \autoref{fig:ocel} b). The corresponding process execution, a sequence, is depicted in the upper right of \autoref{fig:p_exec}. The concurrency information from the graph-based process execution is lost when flattening to a sequence.

\subsection{Object-Centric Process Models}
A process can be represented as a process model, typically a Petri net. Object-centric processes are represented using an object-centric Petri net~\cite{DBLP:journals/fuin/AalstB20}, a Petri net with places of different types and variable arcs which are able to consume more than one token.
\begin{definition}[Object-Centric Petri Net]
An object-centric Petri net is a tuple $\mi{OCPN} = (P,\pi_\mi{pt}, F_\mi{var})$ of 
\begin{itemize}
    \item[$\bullet$] a Petri net $P = (T,P,F,l)$ consisting of transitions $T$, places $P$, arcs $F \subseteq (T \times P) \cup (P \times T)$, and a labeling function $l:T\nrightarrow \mc{A}$,
    \item[$\bullet$] a type function mapping each place to an object type $\pi_\mi{pt}:P\rightarrow \mc{OT}$, and
    \item[$\bullet$] a set of variable arcs $F_\mi{var}\subseteq F$.
\end{itemize}
We define the following notations for object-centric Petri nets:\begin{itemize}
     \item[$\bullet$]  the preset of a transition ${\bullet} t {=} \{p {\in} P \mid (p,t) {\in} F\}$ for $t\in T$.
     \item[$\bullet$] the postset of a transition $t {\bullet}  {=} \{p {\in} P \mid (t,p) {\in} F\}$ for $t\in T$.
     \item[$\bullet$] $tpl(t) {=} \{\pi_{pt}(p) \mid p {\in} {\bullet} t {\cup} t {\bullet}\}$ are the object types associated to transition $t\in T$.
     \item[$\bullet$] $tpl_{nv}(t) {=} \{\pi_{pt}(p) {\mid} p {\in} P \wedge \{(p,t),(t,p)\} {\cap} (F{\setminus} F_{\textit{var}}) {\neq} \emptyset\}$ are the object types with non-variable arcs associated to transition $t\in T$.
\end{itemize}
\end{definition}
An example of an object-centric Petri net is depicted in the lower left of \autoref{fig:p_exec}. Places of different types are depicted with different colors, variable arcs are depicted with double lines. Please note that we dropped the output places of an object type's last transition for presentation purposes. 

Analogously to object-centric and traditional event logs, a traditional Petri net is a special case of an object-centric Petri net where all places have the same type and no variable arcs exist.
The bottom right of \autoref{fig:p_exec} depicts a traditional Petri net.
\begin{definition}[Traditional Petri Net]
An object-centric Petri net  $\mi{OCPN} = (P,\pi_\mi{pt}, F_\mi{var})$ is a traditional Petri net iff $|\mi{range}(\pi_\mi{pt})| = 1 \wedge F_\mi{var} = \emptyset$.
\end{definition}
We describe the state of an object-centric Petri net with a marking.
\begin{definition}[Marking]
Let  $\mi{OCPN} = ((T,P,F,l),\pi_\mi{pt}, F_\mi{var})$ be an object-centric Petri net. A token associates an object with a place. The set of tokens is define by $Q_\mi{OCPN} = \{(p,o) \in P \times \mc{O}\mid \pi_\mi{pt}(p) = \pi_\mi{type}(o)\}$. A marking is a multiset of tokens $M\in \mc{B}(Q_\mi{OCPN})$.
\end{definition}
Given a marking, a transition can be enabled. If it is enabled, it can be executed. The execution of a transition is bound to objects. These are consumed in the input places and produced in the output places. 
\begin{definition}[Binding Execution]
Let  $\mi{OCPN} = ((T,P,F,l),\pi_\mi{pt}, F_\mi{var})$ be an object-centric Petri net.
$B_\mi{OCPN} {=} \{(t,b) \in T {\times} (\mc{OT} {\nrightarrow} \mathcal{P}(\mc{O})) \mid  dom(b) {=} tpl(t) \wedge \forall_{ot {\in} tpl_{nv}(t)} |b(ot)|{=}1\}$ defines the set of all possible bindings. The consumed tokens of a binding $(t,b) {\in} B_\mi{OCPN}$ are defined by $cons(t,b) {=} [(p,o){\in} Q_{\mathit{OCPN}} \mid p {\in} {\bullet} t \wedge o {\in} b(\pi_\mi{pt}(p)) ]$ and the produced tokens are defined by $prod(t,b) {=} [(p,o){\in} Q_{\mathit{OCPN}} \mid p {\in} t {\bullet}  \wedge o {\in} b(\pi_\mi{pt}(p)) ]$. A binding $(t,b) {\in} B$ in marking $M \in \mc{B}(Q_\mi{OCPN})$ is enabled if $cons(t,b) \leq M$. Executing an enabled binding in $M$ leads to marking $M' {=} M - cons(t,b) + prod(t,b)$. Executing an enabled binding $(t,b)\in B_\mi{OCPN}$ in marking $M$ is denoted with $M {\xrightarrow{(t,b)}}M'$. Multiple subsequent enabled bindings can be encoded as a binding sequence  $\sigma {=} \langle (t_1,b_1),\ldots,(t_n,b_n)\rangle {\in} B_\mi{OCPN}^*$. A sequence that starts in marking $M_0$ and results in marking $M_n$ is encoded as $M_0 {\xrightarrow[]{\sigma}}M_n$.
\end{definition}

To define a process model that allows certain behavior, we need start and end points. These are sets of marking for an object-centric Petri net.
\begin{definition}[Accepting Object-Centric Petri Net]
Let  $\mi{OCPN} = ((T, P,\allowbreak  F, \allowbreak l),\pi_\mi{pt}, F_\mi{var})$ be an object-centric Petri net and let $M_\mi{init}, M_\mi{final}\subseteq \mc{B}(Q_\mi{OCPN})$ be initial and final markings of the object-centric Petri net.  $\mi{OCPN}_A = (\mi{OCPN},\allowbreak M_\mi{init}, M_\mi{final})$ is an accepting object-centric Petri net.
\end{definition}

All binding sequences that lead from an initial marking to a final marking define the possible end-to-end behavior of the accepting object-centric Petri net. We define the language of the accepting object-centric Petri net as the set of all possible visible loop-free label sequences in the end-to-end behavior of the object-centric Petri net. In this way, we exclude binding sequences with loops that would render the language of the model infinite.
\begin{definition}[Petri Net Language]
Let $\mi{OCPN}_A = (\mi{OCPN}, M_\mi{init}, M_\mi{final})$ be an accepting object-centric Petri net. All loop-free valid binding sequences of the object-centric Petri net are given by $S(\mi{OCPN}_A) = \{\langle (t_1,b_1),\ldots,(t_n,b_n) \rangle \in B_\mi{OCPN}^* \mid \exists_{M_i \in M_\mi{init}}\; \exists_{M_f \in  M_\mi{final}}\;   M_i {\xrightarrow{(t_1,b_1)}} \cdots {\xrightarrow{(t_n,b_n)}} M_f \wedge \forall_{i,j\in \{1,\ldots,n\} \wedge i\neq j} M_i \neq M_j\}$. The language of the object-centric Petri net is the set of visible transition firing sequences $\Sigma(OCPN_A) = \{\langle  l(t_1),\ldots,l(t_n)\rangle \mid \langle  (t_1,b_1),\ldots,(t_n,b_n)\rangle \in S(\mi{OCPN}_A)  \}$
\end{definition}

An accepting object-centric Petri net can be discovered from an OCEL using the general approach introduced by van der Aalst and Berti~\cite{DBLP:journals/fuin/AalstB20}. A Petri net is discovered for each object type. Given the cardinalities of different objects, the individual Petri nets are merged into one object-centric Petri net, connecting the individual nets at interconnecting transitions (transitions including multiple object types). Place types and variable arcs are assigned according to the types and cardinalities stemming from the individual subnets.
\begin{definition}[Process Discovery]
Let $L=(E, \allowbreak OT \allowbreak ,O,\pi_\mi{act},\pi_\mi{time},\pi_\mi{trace})$ be an object-centric event log. A process discovery algorithm $d(L)= \mi{OCPN}_A$ returns an accepting object-centric Petri net for the object-centric event log.
\end{definition}
This general approach discovers a traditional Petri net if a traditional event log is the input.
The object-centric Petri nets in \autoref{fig:p_exec} are discovered from the OCEL and the flattened event log of \autoref{fig:ocel}.

\subsection{Inter- and Intra-Object Complexity}
In this section, we formally define the inter- and intra-object complexity which were already informally introduced in the introduction. We later use these dimensions in our experimental framework to differentiate which process characteristics and, therefore, which real-life processes benefit most from employing object-centric discovery.
\begin{definition}[Inter-Object Complexity]
Let $\mi{OCPN}_A = (((T,P,F,l),\pi_\mi{pt},\allowbreak F_\mi{var}), M_\mi{init},\allowbreak M_\mi{final})$ be an accepting object-centric Petri net. We define the following model attributes:
\begin{itemize}
    \item[$\bullet$] $\mi{numt}(\mi{OCPN}_A) = |\{t\in T \mid t \in \mi{dom}(l)\}|$ the number of non-silent transitions,
    \item[$\bullet$] $\mi{numot}(\mi{OCPN}_A) = |\mi{range}(\pi_\mi{pt})|$ the number of object types, and
    \item[$\bullet$] $\mi{inter}(\mi{OCPN}_A) = \frac{1}{\mathit{max}(1,|OT|-1)}\frac{\sum_{t\in T} |\{\pi_\mi{pt}(p) \mid p \in \bullet t\}|-1 }{|T|}$  is the inter-object complexity of the model.
\end{itemize}
\end{definition}
The inter-object complexity is defined as the amount of shared transitions between objects. The more transitions are shared between objects, the more the object's control flows are interwoven with each other. Based on the definition, we make two observations.
\begin{observation}
    High inter-object complexities point to many shared transitions. In the extreme case, all objects share all transitions. This would make objects redundant, as they all describe the same control flow. Only one object is necessary to encode the control flow, i.e., a process with an inter-object complexity of 1 is equivalent to a traditional process described by a Petri net. 
\end{observation}
\begin{observation}
    A traditional process has an inter-object complexity of 1, as the single object of the process is involved in all transitions.
\end{observation}
The inter-object complexity only describes the complexity of control flow between objects. To fully capture the complexity of the control flow, we further define the intra-object-complexity, capturing the control flow within objects.
\begin{definition}[Intra-Object Complexity]
Let $\mi{OCPN}_A  = (((T,P,F,l),\allowbreak \pi_\mi{pt},\allowbreak F_\mi{var}),\allowbreak M_\mi{init},\allowbreak M_\mi{final})$ be an accepting object-centric Petri net. For an object type $\mi{ot}\in \mi{range}(\pi_\mi{pt})$, we retrieve the subnet $\mi{OCPN}_{A,\mi{ot}} = (P_\mi{ot},T_\mi{ot},F_\mi{ot},l_\mi{ot})$ with 
\begin{itemize}
    \item[$\bullet$] $P_\mi{ot} = \{p\in P\mid \pi_\mi{pt}(p)=\mi{ot}\}$,
    \item[$\bullet$] $T_\mi{ot} = \{t\in T\mid \exists_{p\in \bullet t \cup t\bullet}\; \pi_\mi{pt}(p)=\mi{ot}  \}$,
    \item[$\bullet$] $F_\mi{ot} = \{(s,t)\in F \mid (s\in P_\mi{ot} \wedge t \in T_\mi{ot} ) \vee (s\in T_\mi{ot} \wedge t \in P_\mi{ot}) \}$, and
    \item[$\bullet$] $l_\mi{ot}(t) = l(t)$ for $t \in T_\mi{ot}$.
\end{itemize}
We define the intra-object complexity for an individual type as $\mi{tioc}(\mi{OCPN}_{A,\mi{ot}}) = \frac{|\Sigma(\mi{OCPN}_{A,\mi{ot}})|}{\mi{numt}(\mi{OCPN}_{A,\mi{ot}})!}$. The intra-object complexity of the model is defined as the average intra-object complexity of all object types $\mi{intra}(\mi{OCPN}_{A}) = \frac{1}{\mi{numot}(\mi{OCPN}_{A})}\cdot\sum_{\mi{ot} \in \mi{range}(\pi_\mi{pt})}\mi{tioc}(\mi{OCPN}_{A,\mi{ot}})$.
\end{definition}
The intra-object complexity quantifies the objects' average deviation from a completely linear control flow. An intra-object complexity of 0 means that all objects follow a strictly linear control flow, while an intra-object complexity of 1 means that all objects would have a completely concurrent control flow.

\section{Experimental Framework}
\label{sec:method}
\begin{figure}[t]
    \centering
    \includegraphics[width=0.6\textwidth]{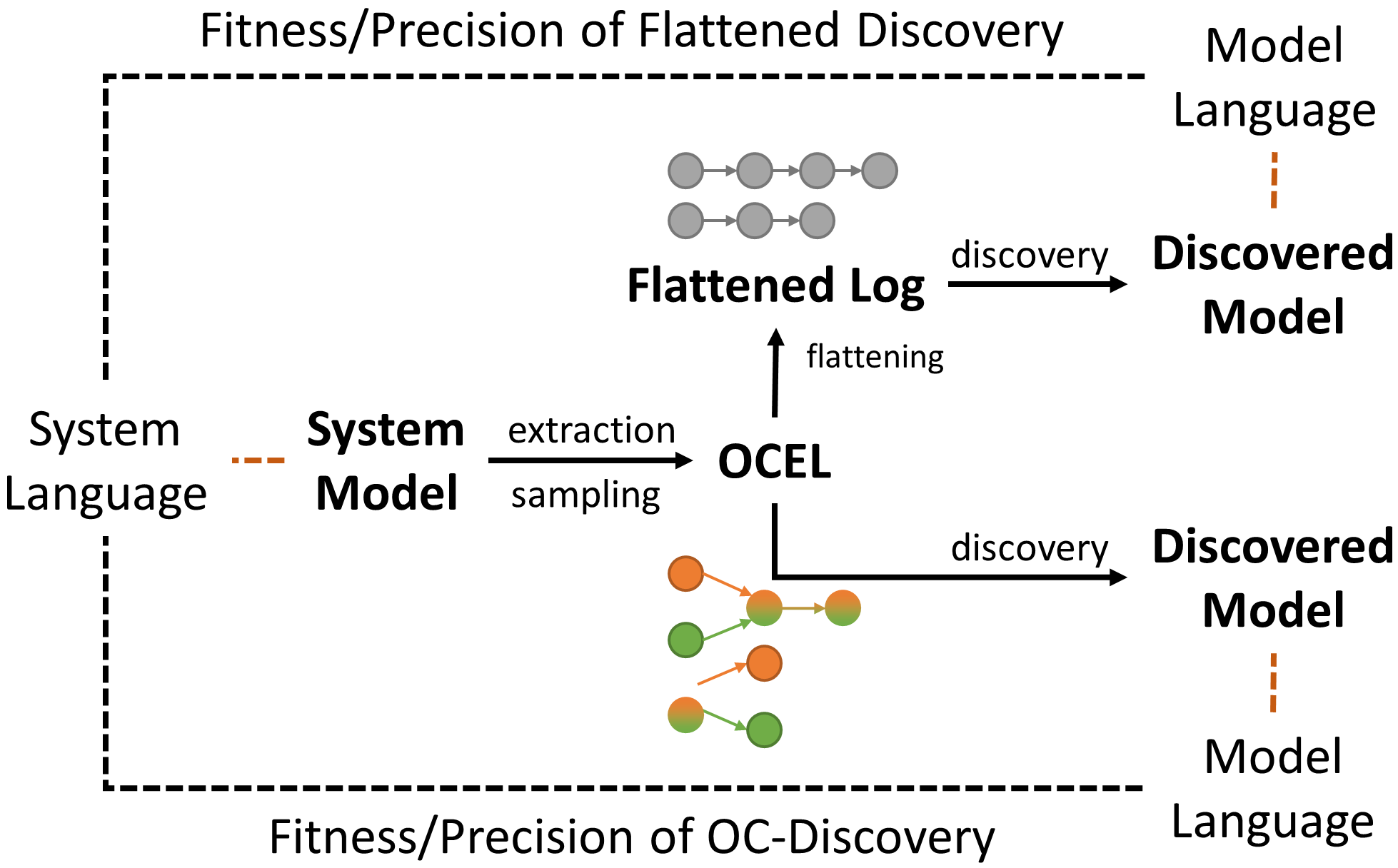}
    \caption{Experimental framework. We generate many different system models with varying characteristics.}
    \label{fig:ex_setup}
\end{figure}
This section introduces the experimental setup to compare process discovery on object-centric and flattened event data which is depicted in \autoref{fig:ex_setup}. The experimental framework consists of three steps: Model generation, event log sampling, and discovery quality assessment.

\paragraph{Model Generation}
\begin{figure}[t]
    \centering
    \includegraphics[width=\textwidth]{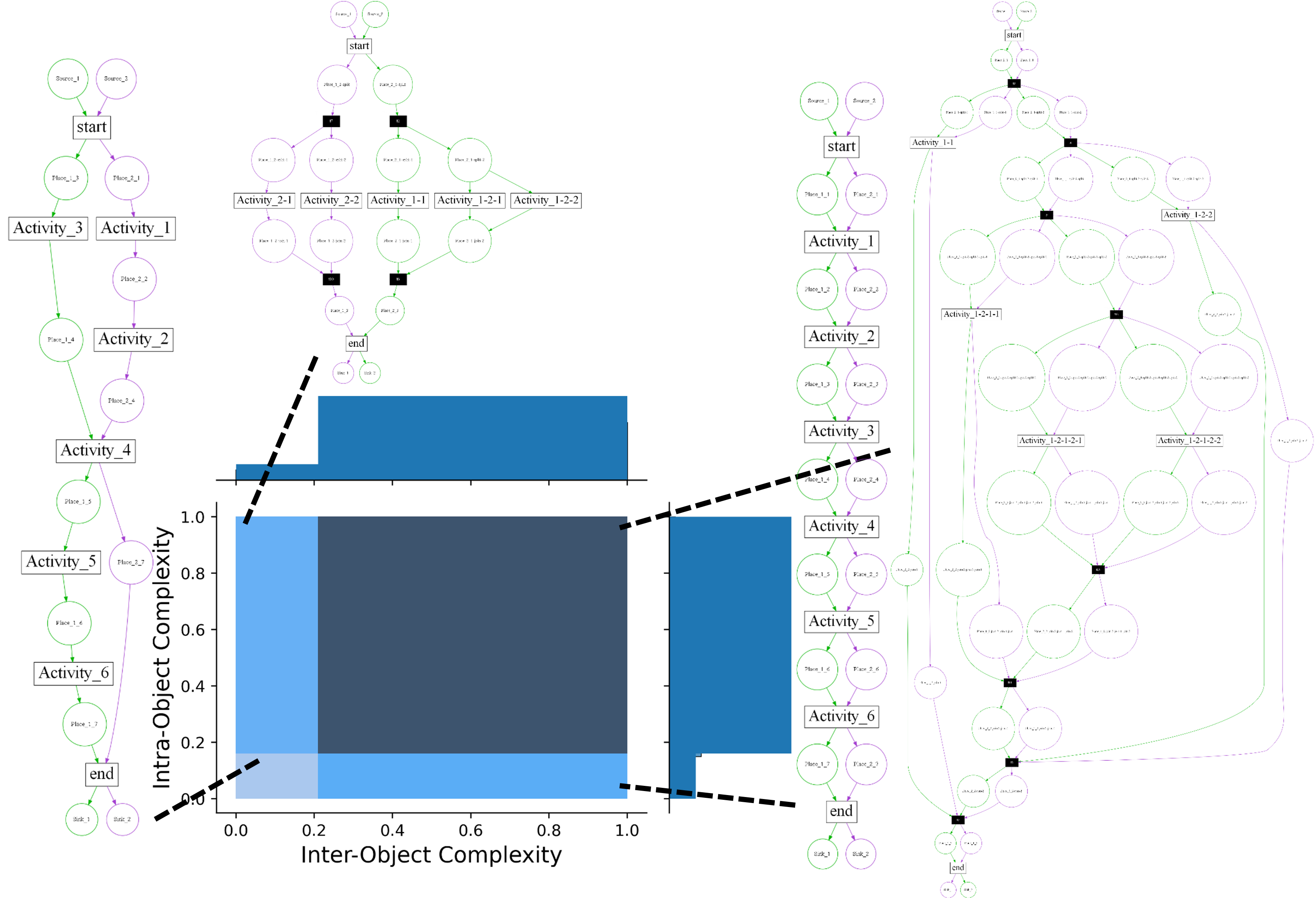}
    \caption{Distribution of generated model characteristics and example models for different combinations.}
    \label{fig:model_char}
\end{figure}
We use a process model called the system model resembling a ground truth process. To cover a wide range of combinations of inter- and intra-object complexity, we generate a large set of different models. 
We depict the distribution of our generated system model along with some exemplary models in \autoref{fig:model_char}. In total, we have randomly generated 44570 system models with 6-8 visible activities (additionally, they can have silent transitions to model AND constructs) and two object types. These models do not have loops, as this would conflict with the next part of the framework, the event log sampling. We address this limitation in the threats to validity section at the end of \autoref{sec:results}. The model sizes are limited to maximum 8 visible transitions, as larger models would produce so many possible traces that an exact computation of the model language, as described in the next section, would become infeasible. However, we complement the experiments with a case study on a large production process, showing the applicability and improvements of discovery also for very large models.

\paragraph{Event Log Sampling}
We extract an event log by sampling from the language of the system model. To do so, we compute the language of the system model, i.e., all possible process executions. We sample a subset of this language to simulate extracting an event log that only contains a part of the possible behavior. 
\begin{definition}[Event Log Extraction]
Let $\mi{OCPN}_A {=} (\mi{OCPN}, M_\mi{init}, \allowbreak M_\mi{final})$ be an accepting object-centric Petri net and $s\in [0,1]$ be an extraction sampling rate. $\mi{sample}(\mi{OCPN}_A,s)\subseteq  \Sigma(OCPN_A)$ samples a subset of the Petri net language corresponding to the sampling rate $s$. The language subset is mapped to an OCEL by $\mi{gen}(\mi{sample}(\mi{OCPN}_A,s),\mi{OCPN}_A) = (E, \allowbreak O \allowbreak ,OT,\pi_\mi{act},\pi_\mi{time},\pi_\mi{trace})$.
\end{definition}
The presented approach allows us to quantify a sampling rate, i.e., how much of the possible model behavior is contained in the event log. Since we computed the full system model language, we can also compare the language of a discovered model to the language of a system model, quantifying how well the discovered model corresponds to the ground-truth system model given the sampling rate. This is explained in the next paragraph
\paragraph{Quantifying Discovered Model Quality}
We compare the quality of discovered process models from the object-centric and flattened event logs. Starting from the OCEL, we discover a model before and after flattening the event log using Inductive Miner~\cite{DBLP:conf/apn/LeemansFA13}. The model quality is defined as the recall/precision of the discovered model language with respect to the language of the system model. When the model is discovered from a sampled event log of the system model, the fitness defined here also measures the generalization capabilities of the discovery technique~\cite{DBLP:conf/caise/PolyvyanyyMG22}.
\begin{definition}[Model Quality]
Let $\mi{OCPN}_A = (\mi{OCPN}, M_\mi{init}, M_\mi{final})$ be a system model and $s\in [0,1] $ be a sampling rate. The sampled OCEL is given by $L = d(\mi{gen}(\mi{sample}(\allowbreak \mi{OCPN}_A,s),\mi{OCPN}_A)$.
The discovered model on the object-centric event log is denoted with $\mi{OCPN}^d_A= d(L)$ and the discovered model on the flattened event log is denoted with $\mi{OCPN}^{\mi{flat},d}_A= d(L^\mi{flat})$.
For any of the two discovered models $\mi{PN}\in\{\mi{OCPN}^{d},\mi{OCPN}^{\mi{flat},d}\}$, we define the fitness $\mi{fit}(\mi{OCPN}_A, \allowbreak \mi{PN}) = \allowbreak \frac{|\Sigma(\mi{OCPN}_A)\cap \Sigma(\mi{PN})|}{|\Sigma_d(\mi{OCPN}_A)|}$ and the precision $\mi{prec}(\mi{OCPN}_A, \allowbreak \mi{PN}) = \frac{|\Sigma(\mi{OCPN}_A)\cap \Sigma(\mi{PN})|}{|\Sigma(\mi{PN})|}$.
\end{definition} 
Using this setup, we retrieve the discovered model quality for traditional and object-centric process discovery for different combinations of inter-object complexity, intra-object complexity, and sample sizes.

\section{Experimental Results}
\label{sec:results}
\begin{figure}[t]
    \centering
    \includegraphics[width=\textwidth]{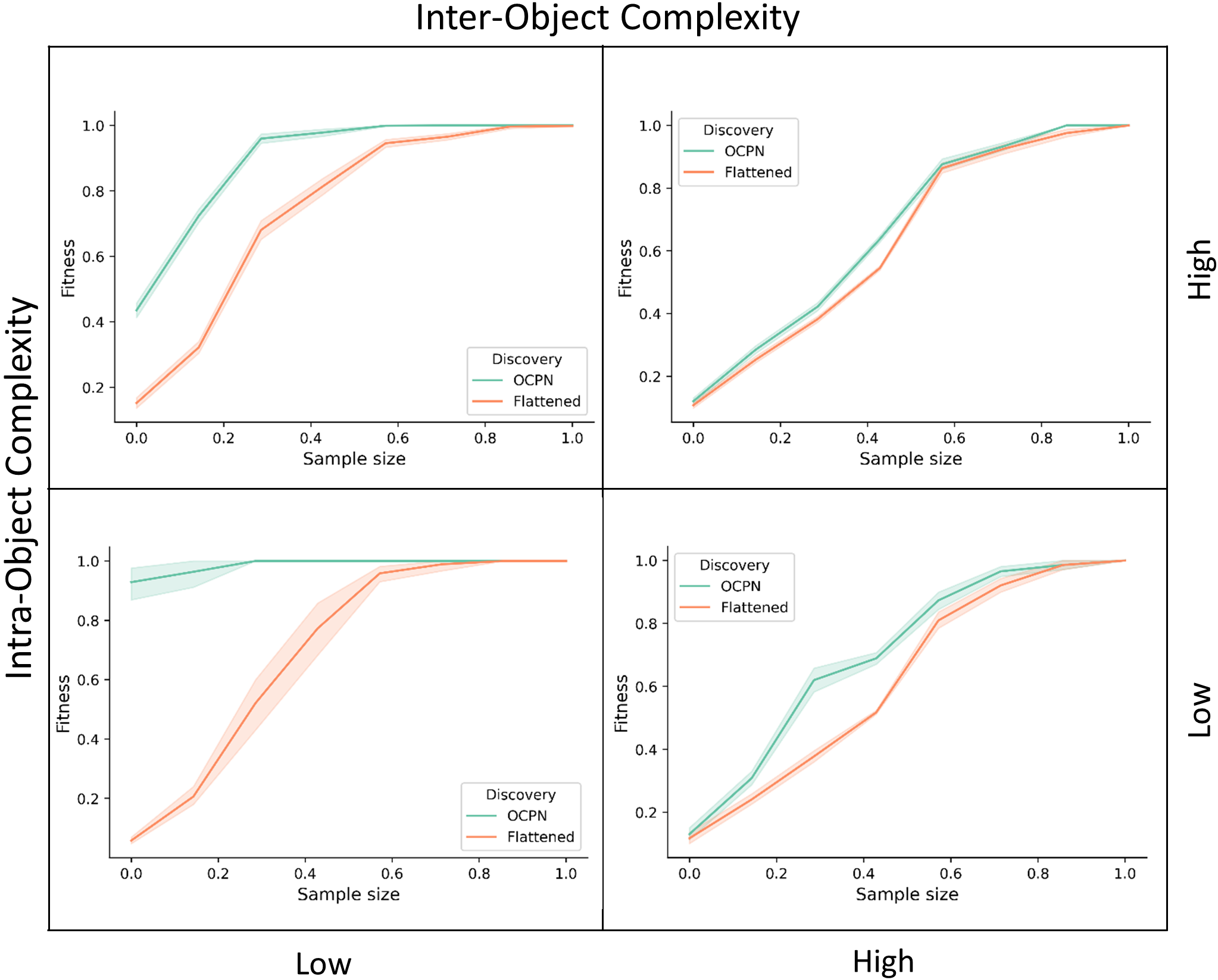}
    \caption{Discovered model fitness depending on the sample size of the event log for the four different quadrants of our process taxonomy.}
    \label{fig:results}
\end{figure}

We present our experimental results in this section. The implementation is publicly available\footnote{\url{https://github.com/jn-adams/ObjectCentricDiscovery}}. We present the dependency between discovered model quality and event log sample size for traditional and object-centric discovery on system models with varying inter- and intra-object complexity, as described in \autoref{sec:method}. The results are depicted in \autoref{fig:results} (fitness) and \autoref{fig:results_precision} (precision).

{
\setlength{\belowcaptionskip}{-10pt}
\begin{figure*}[t]
        \centering
        \begin{subfigure}[b]{0.49\textwidth}   
            \centering 
            \includegraphics[width=\textwidth]{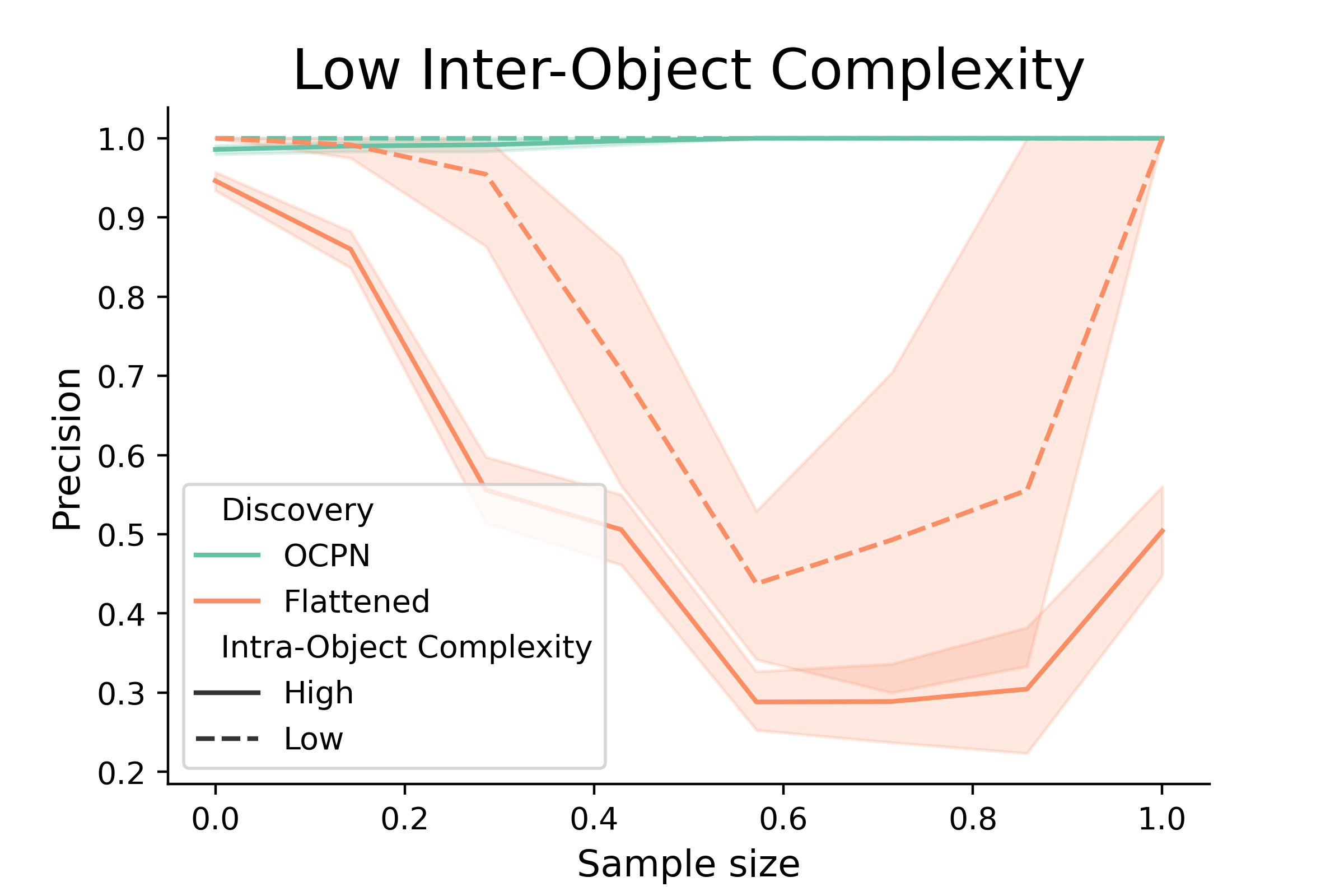}
               
            \label{fig:res3}
        \end{subfigure}
        \hfill
        \begin{subfigure}[b]{0.49\textwidth}   
            \centering 
            \includegraphics[width=\textwidth]{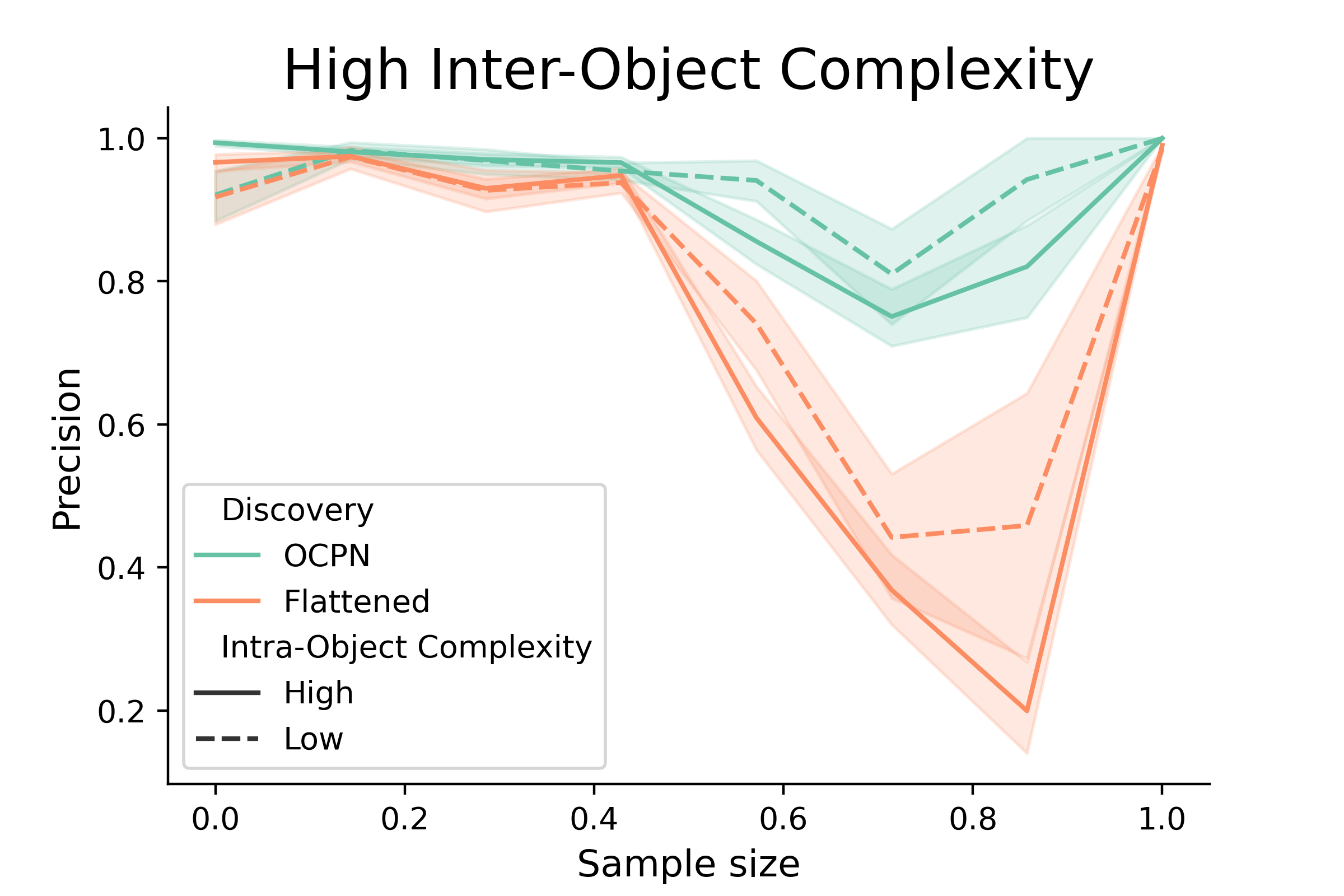}
                
            \label{fig:res4}
        \end{subfigure}
        \caption{Discovered model precision depending on the sample size of the event log for the four different quadrants of our process taxonomy.} 
        \label{fig:results_precision}
    \end{figure*}}
We split the models according to their characteristics using low/high inter- and intra-object complexity. The decision for low and high values are made according to \autoref{fig:model_char}, i.e., an inter-object complexity of more than 0.2 is considered high and an intra-object complexity of less than 0.15 is considered to be low. 
We choose both values for the following reasons: First, high inter-object complexity indicates high redundancies between object types, i.e., the system model is behaviorally very close to a traditional Petri net. By choosing a low threshold we single out system models that are significantly different from traditional process models. Second, high intra-object complexity models indicate a lack of structure in the process, i.e., no order is enforced. By choosing a low threshold, we can single out system models that show high degrees of sequentiality in one bin and less structured models in the other bin.

The highest quality differences can be observed when the system model exhibits low levels of inter-object complexity, especially when the intra-object complexity is also low, i.e., each object has a relatively sequential path and only some interaction points with other objects. In this case, the average fitness of the discovered process model for extracted OCELs with a low sample rate is already 0.9. The fitness of the discovered model from the flattened event log is very low, starting at almost 0. Mapping this back to the initial taxonomy depicted in \autoref{fig:taxonomy}, object-centric discovery can significantly reduce the data required for manufacturing processes, supply chains, and composite workflows. Especially for processes with relatively sequential subprocesses, like assembly lines, object-centric discovery allows for discovery with much less data compared to traditional discovery. This means, that discovery for larger models, which was prohibitive due to the data requirements in traditional discovery, is now enabled using object-centric discovery. We will also see this in the case study in \autoref{sec:casestudy}. Models with high inter-object complexity cannot benefit that much from switching to object-centric discovery. Since high inter-object complexity points to redundant control flows and is, in the extreme case, equivalent to traditional process models, these results are not surprising. It is notable, that -- within this evaluation -- object-centric discovery does not worsen results.

When considering the precision of discovered models, the quality differences are larger for models with high intra-object complexity, although precision levels for low intra-object complexity are, generally, higher. This can be explained by low levels of intra-object complexity allowing more concurrent behavior, such that the discovery of flower-like models will be more precise than for more restrictive system models. In general, we observe that the high amounts of sequences produced by concurrent behavior push the discovery algorithm to discover flower models for the flattened event logs, significantly reducing the precision.

\subsection{Threats to Validity}
As our experimental framework computes the exact languages of the models, it is computationally quite demanding, leading us to restrict the system model generation in two major ways: without loops and only limited to eight visible transitions. These are also the main limitations of this experimental framework: First, our results cannot be generalized to process models containing many loops. Second, our generated models only cover relatively small models of 8 visible transitions. While the results are already very clear and the difference should only increase for larger models, our experiments cannot prove this claim. Therefore, we complement the experiments with a case study performing object-centric discovery on a large-scale production process and comparing the results to traditional process discovery. We aim to mitigate the limitations of our experimental framework using this case study, as it shows that object-centric discovery is feasible for large processes and provides better results than traditional discovery.

\section{Case Study}
\label{sec:casestudy}
{
\setlength{\belowcaptionskip}{-12pt}
\begin{figure}[t]
    \centering
    \includegraphics[width = 0.95\textwidth]{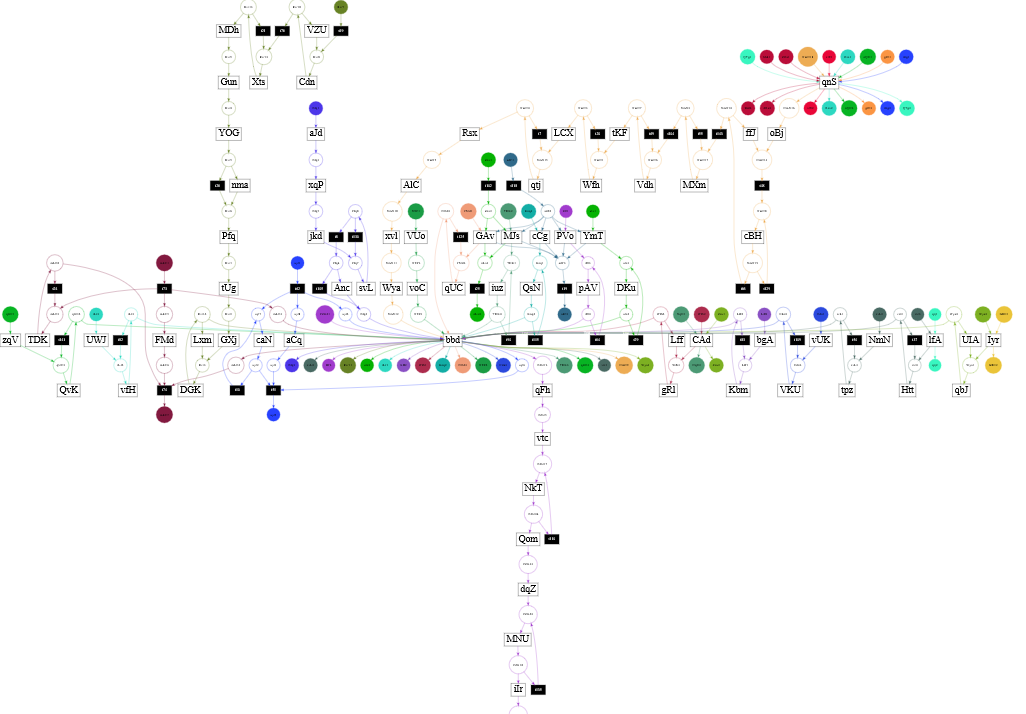}
    \caption{Discovered process model of a production process using an OCEL. Different components follow their individual paths and are assembled into a merged component. }
    \label{fig:ocpn_disc}
\end{figure}}
We use object-centric event data from a production process in the  German manufacturing company Heidelberger Druckmaschinen AG~\cite{DBLP:conf/icpm/BrockhoffUTGA21}. The confidential original event log contains hundreds of process executions with more than 800 activities. Object-centric discovery can be applied to the original event log, however, the resulting visualization is too large for human comprehension. Therefore, we limit our analysis to a subprocess of 105 activities. Our sublog contains 13 process executions. We, first, apply object-centric process discovery and, second, flatten the event log and apply traditional process discovery. Even though we do not know the true ground truth process to provide an exact sample rate, it would be close to 0 given the large number of concurrent activities and the low number of process executions. The results are depicted in \autoref{fig:ocpn_disc} and \autoref{fig:flat_disc}.

The structure of the process can be rediscovered using object-centric discovery: Individual concurrent object paths are visible, ending up in one assembly activity that combines individual components into an output component. This output component follows its individual path afterward. The low number of process executions is not an issue when using object-centric discovery, it can still rediscover the concurrency between objects. This is not the case for process discovery on flattened event data. The flattened process discovery algorithm shows a flower model, i.e., the structure of the process is completely lost. Furthermore, it is generally infeasible to expect that a flattened event log would contain enough process executions such that concurrency could completely be rediscovered: 100 partly concurrent activities can generate a large number of possible activity sequences, such that the data requirements of traditional process discovery would be beyond any feasible amount of produced products, i.e., the available sample size.

{
\setlength{\belowcaptionskip}{-12pt}
\begin{figure}[t]
    \centering
    \includegraphics[width = 0.95\textwidth]{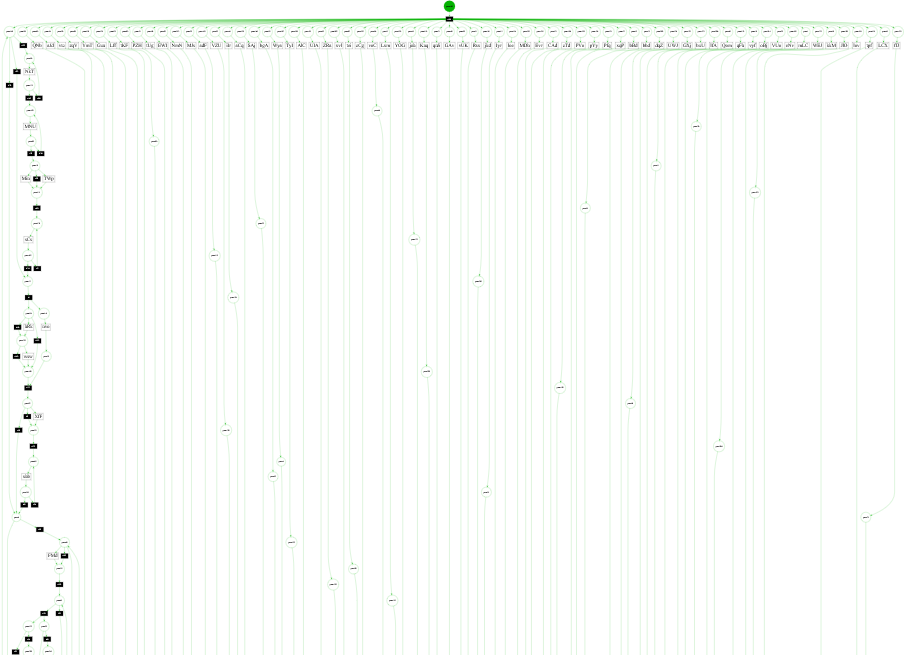}
    \caption{Discovered process model of a production process using the flattened object-centric event log. The process is mostly a flower model and the screenshot only shows part of the width of this flower model.}
    \label{fig:flat_disc}
\end{figure}
}
\section{Conclusion}
\label{sec:conclusion}
In this paper, we investigated the data requirement reduction that object-centric process discovery yields over traditional discovery. We categorize different real-life processes according to control-flow complexity across and within subprocesses, and formally define these dimensions. We define an experimental framework that generates models across these dimensions and assess the reduction in data requirements when using object-centric discovery over traditional discovery. We can show that the data requirements are drastically reduced, especially for low inter-object complexity processes models like production processes or supply chains. To address limitations of the experimental framework w.r.t. the size of generated models we complement our evaluation with a large-scale production process case study.  We show, that object-centric discovery captures the production process much better than traditional process discovery. In this case, the data requirements of traditional discovery would even exceed the number of produced items, rendering traditional process discovery infeasible. Our results have significant implications for the application of process discovery to large-scale processes: Areas once infeasible for discovery due to large data requirements are now accessible using object-centric process mining. This will foster the adaption of process discovery in the future.

\section{Acknowledgments}

We thank the Alexander von Humboldt (AvH) Stiftung for supporting our research. Funded by the Deutsche Forschungsgemeinschaft (DFG, German Research Foundation) under Germany’s Excellence Strategy---EXC-2023 Internet of Production—390621612.
\bibliographystyle{splncs04}
\bibliography{bibliography}
\end{document}